\documentclass{aa}  
\usepackage{txfonts}
\usepackage{graphicx}
\usepackage{natbib}
\usepackage{multirow} 
\usepackage{lscape}
\usepackage{longtable}
\usepackage{amssymb,amsmath}
\usepackage{color}
\bibpunct{(}{)}{;}{a}{}{,} 
\newcommand{\micron}{\mbox{$\mu$m}}

\defcitealias{Dionatos:10a}{Paper I}

\begin{document}

   \title{\emph{Herschel} spectral-line mapping of the HH211 \\ protostellar system$^{\star}$}


   \author{Odysseas Dionatos
          \inst{1}
          \and
          Tom Ray\inst{2}
          \and
          Manuel G\"udel\inst{1}
                   }

   \institute{Institute for Astronomy (IfA), University of Vienna,
              T\"urkenschanzstrasse 17, A-1180 Vienna\\
              \email{odysseas.dionatos@univie.ac.at}
         \and
             Astronomy \& Astrophysics Section, Dublin Institute for Advanced Studies,
          31 Fitzwilliam Place, Dublin 2, Ireland 
             \thanks{\emph{Herschel} is an ESA space observatory with science instruments provided by European-led Principal Investigator consortia and with im- portant participation from NASA.}
             }


 
  \abstract
   {Mid- and far-infrared observations of the environment around embedded protostars reveal a plethora of high excitation molecular and atomic emission lines. A number of different mechanisms for the origin of these lines have been proposed, including shocks induced by protostellar jets and radiation by the embedded protostar interacting with its immediate surroundings.}
   {We employ extended spectral-line maps that spatially resolve regions where diverse excitation processes appear to dominate. Studying the morphology and excitation of the most important molecular and atomic coolants, we aim to constrain the physical conditions around the embedded protostellar system HH\,211-mm.}
   {Spectro-imaging observations with Herschel/PACS provide emission from major molecular (CO, H$_2$O and OH) and atomic coolants (e.g. [\ion{O}{i}]). Emission line maps reveal the morphology of the observed emissions and allow associations between the different species. Comparisons are also made with mid-infrared line-maps from Spitzer and sub-mm interferometers. The excitation conditions of the detected molecular species along with the ortho-to-para ratio of water are assessed through Boltzmann diagrams. Further investigations focus on constraining the CO/H$_2$ ratio in shocks and the mass flux of the atomic jet as traced by [\ion{O}{i}].}
   {Molecular lines are exited mainly at the terminal bowshocks of the outflow and around the position of the protostar. All lines show maxima at the SE bowshock with the exception of water emission that peaks around the central source. Excitation analysis in all positions shows that CO and H$_2$O are mainly thermally excited at $T_{\rm ex}$ $\sim$ 350\,K and 90\,K respectively, with the CO showing a second temperature component at 750\,K towards the SE peak. Excitation analysis breaks down in the case of OH at the blue-shifted bowshock, indicating that the molecule is non-thermally excited. Comparisons between the CO and H$_2$ column densities suggest that the $X[\rm{CO}]$ value in shocks can be up to an order of magnitude lower than the canonical value of 10$^{-4}$. The water ortho-to-para ratio around the protostar is only 0.65, indicating low-temperature water ice formation followed by non-distructive photodesorption from the dust grains. The two-sided total atomic mass flux estimated from the [\ion{O}{i}] jet sums to 1.65\,$\times 10^{-6}$~M$_{\odot}$~yr$^{-1}$, a value that is very close to the mass flux previously estimated for the SiO jet and the H$_2$ outflow.}
   {The bulk of the cooling from CO, OH and [\ion{O}{i}]  is associated with gas excited in outflow shocks, with the blue-shifted (SE) outflow showing evidence of a shock-induced UV field responsible for the OH emission. Water lines around the protostar reveal a very low ortho-to-para ratio that can be interpreted in terms of formation from a primordial gas reservoir in the envelope. Finally comparisons of the [\ion{O}{i}] jet mass-flux to the mass fluxes derived for SiO and H$_2$ renders HH\,211 the first embedded system where an atomic jet is demonstrably shown to possess enough momentum to drive the observed molecular jets and large scale outflows.} 

   \keywords{Stars:formation -- 
                ISM: jets and outflows --
                ISM: kinematics and dynamics --
                ISM: atoms \& molecules --
                ISM: abundances --
                ISM: individual objects: HH211-mm               }

   \maketitle
%

\section{Introduction}

 The initial formation phases of low-mass stars are driven by the interplay of complex processes that allow a rapid growth of the protostellar object. Mass accretion is most of the time accompanied by the launching of high-velocity jets and wide-angle winds that carve up the surrounding envelope forming outflow cavities. At larger distances, protostellar ejecta moving at supersonic velocities  generate shocks that compress and heat-up the surrounding medium \citep{Frank:14a}. At the same time UV radiation produced from the accretion shocks very close to the protostar can readily escape through the outflow cavities, exciting the envelope material \citep[e.g.][]{Visser:12a}. The main cooling mechanism of the gas excited by all these processes is  through the emission of molecular and atomic lines, lying predominantly within the near- to the far-IR window of the electromagnetic spectrum. Backtracking to the exact excitation process through the observed line emissions is however not straightforward. Many processes may act simultaneously, within the same, unresolved region, contributing to the total line emission, which then leads to many possible, often degenerate interpretations \citep[e.g.][]{Visser:12a, Karska:18a}. In this respect, observations that can spatially separate regions where specific excitation processes dominate can provide unique probes to disentangle the precise line-excitation within specific zones around low mass protostars.

Due to its pristine appearance, HH 211 stands out amongst the youngest, best-studied outflows which has made it become a text-book example. Being very young and relatively compact (estimated dynamical timescales are less than 500~yr, see also Sect.~\ref{sec:4.3}), it has a rather simple and well-defined geometry consisting of a central young stellar object, nearly symmetric outflows and two bright terminal shocks at the opposite outflow ends. It is located in the IC\,348 complex in the constellation of Perseus at a distance of $\sim$250 - 320\,pc \citep{Enoch:06a, Lada:06a}, the former value adopted for consistency with previous studies also in this work. The central source has been classified as a low-mass, low-luminosity Class 0 protostar \citep[L$_{\rm bol}\sim$3.6\,L$_{\odot}$,][]{Froebrich:05a}. The area between the position of the protostellar source and the bright, near-IR terminal bow-shocks \citep[ lying at $\sim45\arcsec$,][]{McCaughrean:94a}  is filled with low-velocity (<10\,km\,s$^{-1}$), CO outflow gas \citep{Gueth:99a} and a pair of highly collimated, molecular jets observed in CO and SiO,  reaching line-of-sight velocities of $\sim$40\,km\,s$^{-1}$ \citep[e.g.][]{Gueth:99a, Hirano:06a, Lee:07a}. Detailed observations of the SiO jet  constrained the transverse velocity to 115\,km\,s$^{-1}$ and revealed a reflection-symmetric wiggling pattern that has been attributed to the possible existence of a low mass binary companion closely orbiting the jet-driving source \citep{Lee:10a, Jhan:16a}. The knot spacing corresponds to a period of $\sim20$\,yr, however this cannot be directly related to the period of the binary.
The blue- and red-shifted lobes, pointing to the southeast (SE) and northwest (NW), respectively, are well separated as the inclination of the outflows with respect to the plane of the sky is measured to be $\sim9^{\circ}$ \citep{Jhan:16a}. While the terminal shocks of the two lobes appear rather symmetric in the near-IR, a high-resolution Spitzer map at the tip of the blue lobe revealed strong OH emission of non-thermal origin \citep{Tappe:08a}. Followup Spitzer observations at different positions around the protostellar source and on the red-shifted lobe \citep{Tappe:12a} did not detect highly excited OH emission elsewhere.

Mid-IR slit-scan maps with Spitzer/IRS \citep[][hereafter Paper I]{Dionatos:10a} revealed a ladder of excited rotational H$_2$ transitions along with a series of atomic lines ([\ion{Fe}{ii}],[\ion{Si}{ii}] and [\ion{S}{i}]). H$_2$ emission was found coincident to the CO and SiO jet at distances of $\sim5\arcsec$ from the protostar while atomic lines, likely tracing an atomic jet, were detected very close to the position of the protostellar source. Excitation analysis of the H$_2$ emission revealed two dominant temperature components, tracing ``cool'' and ``warm'' gas at $\sim$300\,K and $\sim$700-1200\,K. The ``cool'' component dominates the outflow mass, showing a factor of 10 higher column densities when compared to the ``warm'' component. The cool H$_2$ mass flux, corrected for possible beam dilution, was found to be  comparable to that inferred from CO observations, assuming a CO abundance of $8.5 \times 10^{-5}$ \citep{Lee:07a}. In addition, the ortho to para ratio for the H$_2$ molecule at the outflow positions was found very close to the thermal equilibrium value of 3. Analysis of the atomic emission suggested that iron and silicon are heavily depleted onto dust grains as the thrust of the atomic jet derived from those lines is apparently not sufficient to drive the large-scale H$_2$ and CO structure. 
  
In this paper we present Herschel/PACS observations covering the entire HH211 bipolar outflow. In terms of wavelength and angular resolution, Herschel maps are complementary to the ones from Spitzer. The emission from molecules such as CO, H$_2$O, OH and [\ion{O}{i}] detected in this work is commonly traced with Herschel around protostellar sources \citep[e.g.,][]{Karska:18a}. The morphological characteristics, however, provided by the extended line maps presented in this work can provide insights into the different excitation mechanisms involved. The paper is organized as follows. In Sect.~\ref{sec:2} we discuss the reduction of the PACS observations and in Sect.~\ref{sec:3} we present the emission maps for all detected lines and discuss basic morphological patterns. In Sect.\ref{sec:4} we perform an excitation analysis for CO, H$_2$O and OH. In addition we determine the CO/H$_2$ in shocks and estimate the ortho-to-para ratio of the H$_2$O lines and discuss possible implications for the thermal processing history of the protostellar material. Finally we derive the atomic mass-loss rate from the  [\ion{O}{i}] jet emission, compare our estimations to the corresponding values for SiO and H$_2$, and discuss the possibility of a jet-powered outflow. Finally in Sect.~\ref{sec:5} we summarize our main conclusions.    
     

   
\section{Observations and data reduction} \label{sec:2}

 Observations were retrieved from the Herschel Science Archive (HSA\footnote{http://archives.esac.esa.int/hsa/whsa/}) and were obtained on the 11th of February 2013 as part of the ``Must-Do'' program executed within the Director's Discretionary Time (DDT). The Photodetector Array Camera and Spectrometer \citep[PACS;][]{Poglitsch:10a} onboard the Herschel Space Observatory  was used in range-scan spectroscopy mode, providing the complete wavelength coverage available with PACS ($\sim$\,55 - 210\,$\mu$m, R $\sim$ 1000 - 3000). Nyquist sampled raster maps spanning $\sim72\arcsec\times63\arcsec$ were obtained for three positions:   
one centered at the protostellar source HH211 ($\alpha_{J2000}$= 03$^h$ 43$^m$ 56$^s$.81
$\delta_{J2000}$=+32$^d$ 00$^m$ 49$^s$.89'') and another two pertaining the positions of the terminal bow-shocks at offsets of $\sim$40$\arcsec$ to the northwest (NW) and southeast (SE) from the protostar. In this configuration there is an overlap of $\sim$\,15$\arcsec$ between consecutive maps which however cover the entire extent of the HH211 outflow. Integration times are $\sim$\,1.32 and 2.44 hrs for the blue and the red modules (short- and long-ward of 100\,$\mu$m), respectively, and for each pointing position which sums up to a total integration time of 11.3 hrs.    

Observations were initially processed with version 15 of the Herschel Interactive Processing Environment (HIPE).  The resulting spaxel scale of the reduced maps is 3$\arcsec$ while the final map dimensions are $\sim140\arcsec\times90\arcsec$. Reduced maps were further processed with home-grown pipelines. Line fluxes were calculated by Gaussian-fitting of emission-line components on each spaxel after removing a first order polynomial baseline. Line-maps were reconstructed by re-projecting the line fluxes measured for each spaxel on a regular grid with a pixel size equal to that of the original cubes. Spectral cubes in the range between 96 and 103 $\micron$ are exceedingly noisy and show irregular variations in the continuum flux levels. Abnormally low flux levels are also evident beyond 190 $\micron$ along with ghost spectral features due to leakage of emission from higher orders. Therefore, line fluxes from these two spectral segments were not included in the analysis that follows.

\begin{figure*}[!ht]
\includegraphics[width=17.9cm]{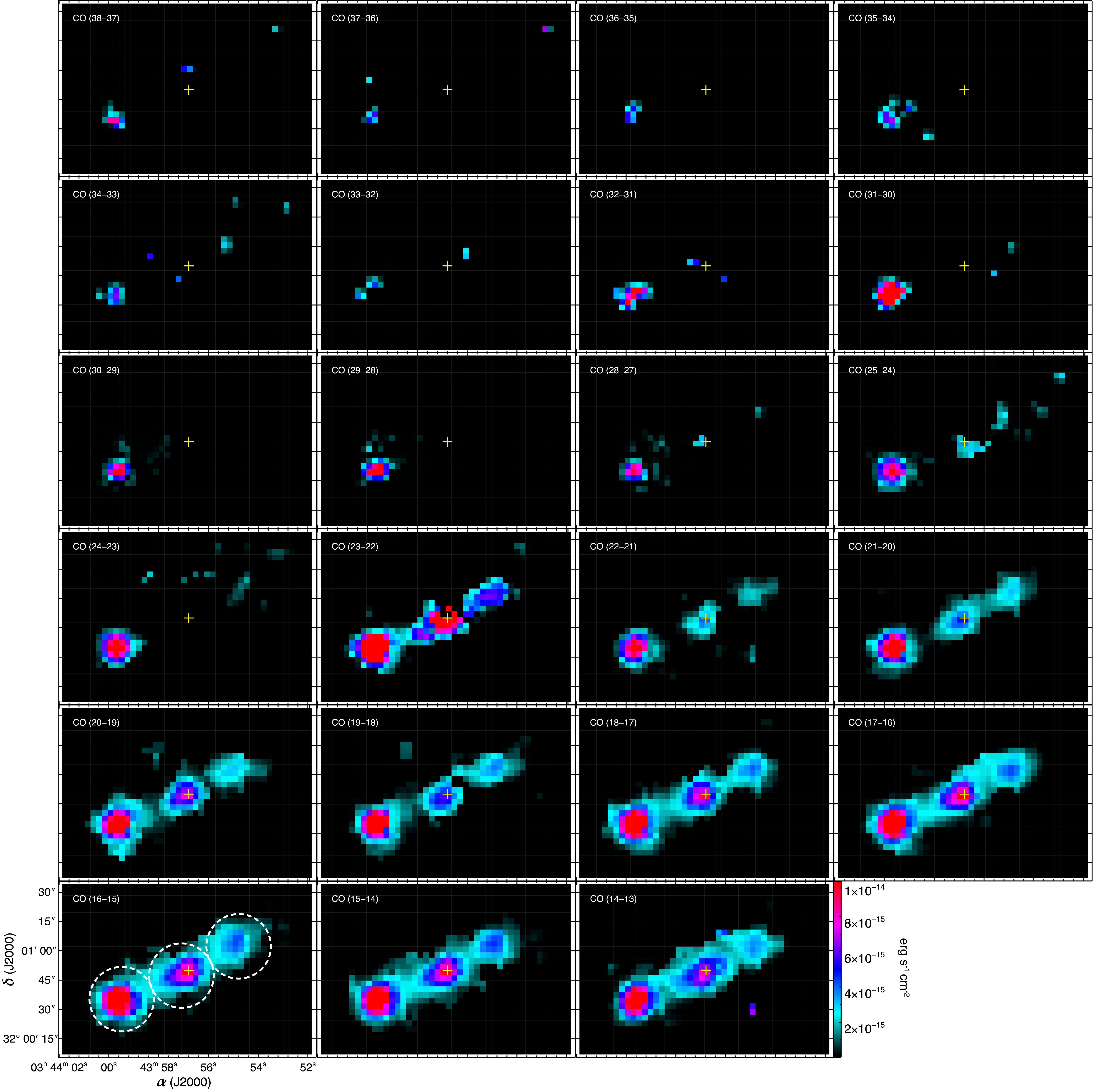}
\caption {CO line maps, starting from the $J = 38 - 37$ transition at 69~$\mu$m in the upper left corner to $J = 14 - 13$ at 183~$\mu$m  at the lower right. The $J = 23 - 22$ map is blended and dominated in the central and north-west regions by the $4_{14}-3_{03}$ o-H$_2$O line. The yellow cross indicates the position of the protostellar source and the dashed circles in the lower left panel delineate the three extraction regions for which excitation diagrams are constructed (see Sect.~\ref{sec:4.1}).} 
\label{fig:1}
\end{figure*}

 \section{Emission line morphology}
 \label{sec:3}

The spectral cubes resulting from the reduction process show strong line emission from CO, H$_2$O, OH and [\ion{O}{i}]. As a general trend line emission appears extended, with the exception of higher energy transitions where it can be rather localized. Line maps from different molecular and atomic tracers display rather diverse morphological patterns. In the following paragraphs we present the reconstructed line-emission maps and discuss their main morphological characteristics.

\begin{figure*}[!ht]
\includegraphics[width=17.9cm]{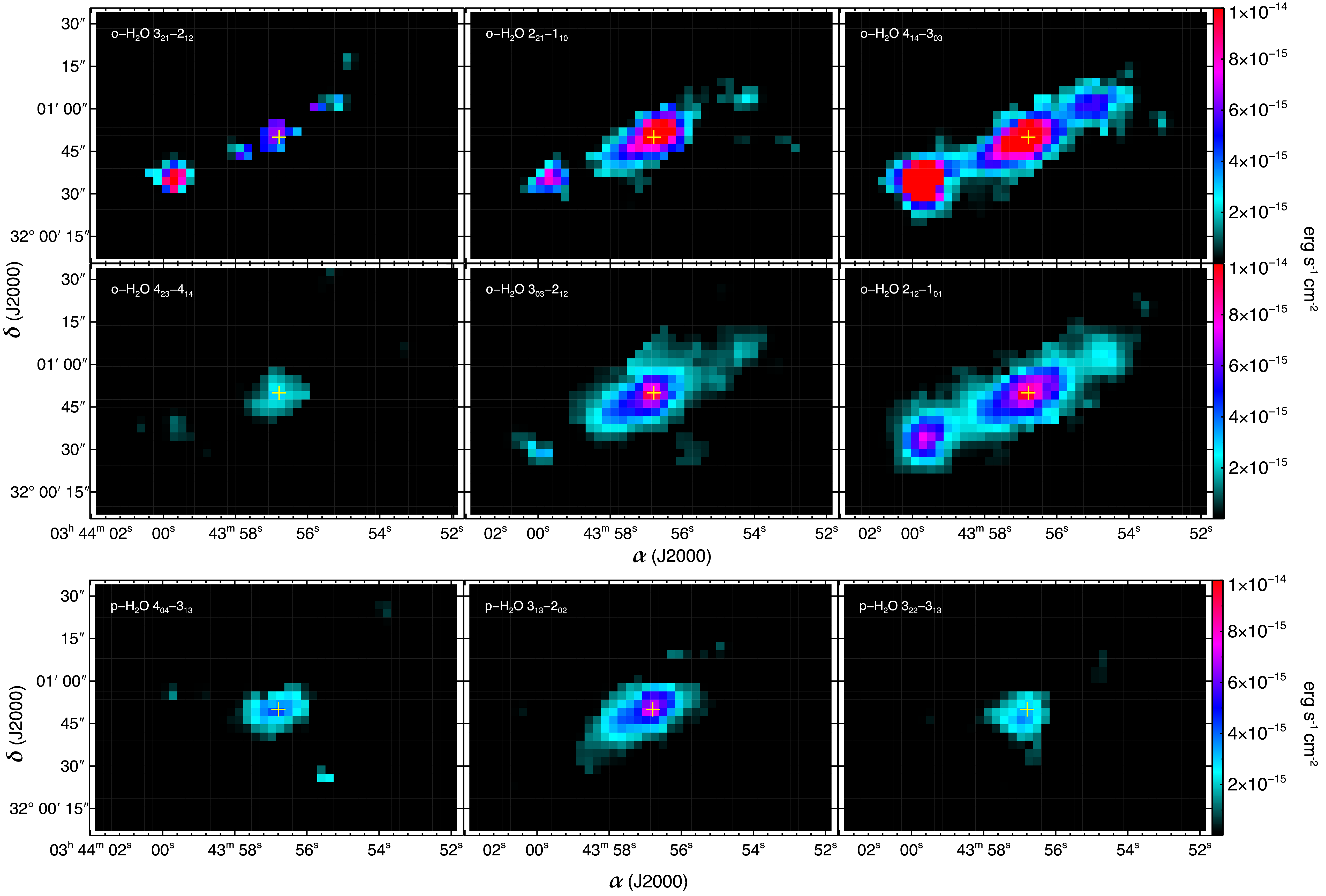}
\caption {Line maps for the ortho and para nuclear spin isomers of H$_2$O (upper and lower panels, respectively). The $4_{14}-3_{03}$ map in the upper right panel is blended with the $J_{\rm up}=23$ CO line. The yellow cross indicates the position of the protostar HH211-mm.} 
\label{fig:2}
\end{figure*}

 
Figure~\ref{fig:1} presents the flux distribution for all detected transitions of carbon monoxide. The highest $J$ transitions observed with blue PACS modules show some persisting noise, however all rendered maps display strong emission from a localized cluster of pixels towards the SE. This cluster becomes brighter in subsequent maps with decreasing $J_{\rm up}$ and remains the brightest region of emission for all CO transitions. Maps with $J_{\rm up}\leq28$ show some emission around the position of the protostellar source HH\,211-mm and towards the tip of the NW lobe, however this emission becomes significant only for the maps with $J_{\rm up}\leq22$.  The CO emission around the protostellar source peaks slightly to the SE from the position of HH\,211-mm and remains brightest, second only to the SE peak in all $J_{\rm up}\leq22$ maps. The NW tip of the outflow becomes apparent in the maps with $J_{\rm up}\leq22$, yet it remains fainter compared to the emission from the SE and on-source positions. This applies also in the $J_{\rm up}=23$ map where CO is blended with the $4_{14}-3_{03}$ o-H$_2$O line and therefore the  morphology corresponds to the combined emission pattern from both molecules (see also Fig.~\ref{fig:2}). The shape of all regions of CO emission appears rather rounded, with the only exception being the maps for $J_{\rm up}\leq16$ at the on-source position that appear more elongated and closer to the water emission pattern (Fig.~\ref{fig:2}). The CO emission trend from all transitions observed with Herschel/PACS appears quite different compared to the morphology of lower-$J$ (2-1 and 3-2) CO maps obtained with interferometers \citep{Gueth:99a, Lee:07a}, a fact that cannot only be attributed to the difference in angular resolution (see also the [\ion{O}{i}] maps below). High-$J$ CO emission appears mostly localized at the positions of the terminal bowshocks and around the protostar and does not seem to trace the outflow-lobe structure or the molecular jets, as observed in low-$J$ CO lines. The $J_{\rm up}\leq22$ pattern resembles the H$_2$ and [\ion{Si}{ii}] emission presented in \citetalias{Dionatos:10a}, with the difference that mid-IR line emission close to the protostar is highly extincted by the surrounding envelope.

 \begin{figure*}[!ht]
\includegraphics[width=17.9cm]{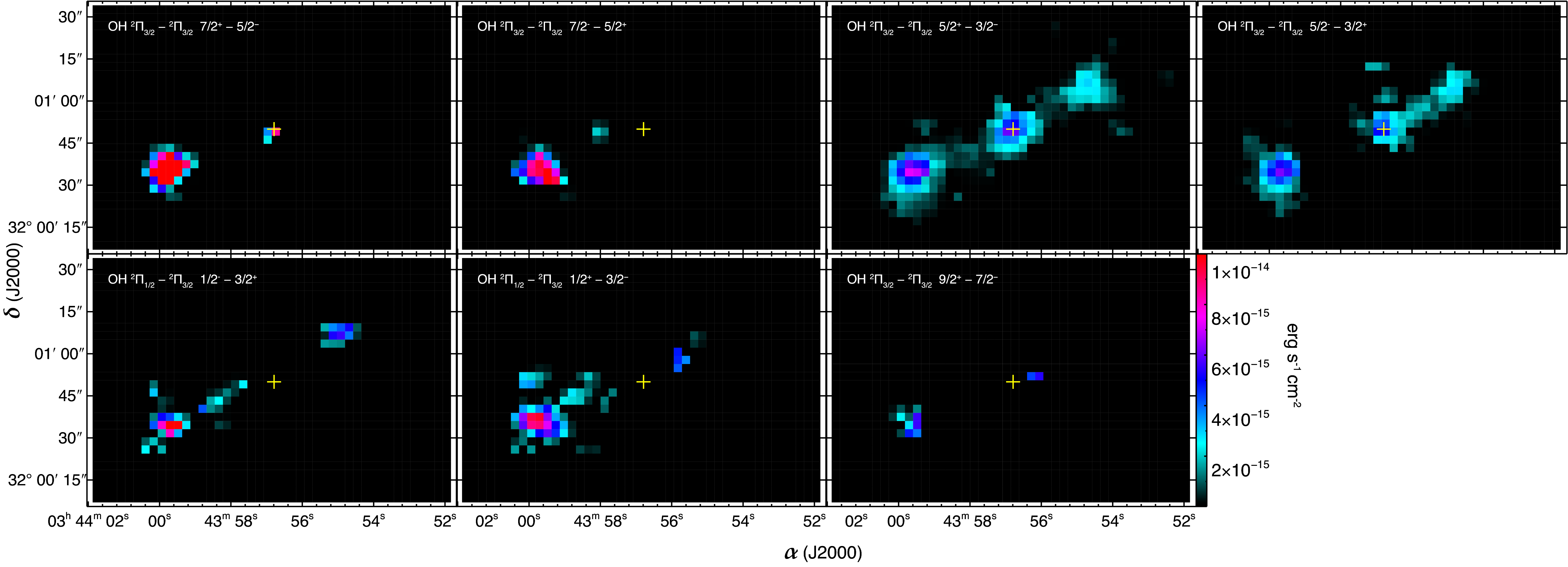}
\caption {Line maps for the resolved OH line doublets. The yellow cross indicates the position of the protostar HH211-mm.} 
\label{fig:3}
\end{figure*}

In Fig.~\ref{fig:2}, we present the emission pattern of the ortho- and para- forms of the water molecule (upper and lower panels, respectively). Water appears to be, in most cases, centralized around the protostar. Ortho transitions often display some weaker emission toward the SE and NW positions, with the SE emission being in most cases stronger in comparison to the NW. The most prominent map showing H$_2$O emission in the outflow is from the $2_{12}-1_{01}$ transition at 179.5$\micron$. Maps of the same transition observed with PACS in line-spectroscopy mode \citep{Tafalla:13a} are in excellent agreement to the ones presented here. In contrast to CO, the shapes at the maxima in the ortho-water emission maps are ellipsoidal, with the major axis extending towards the SE-NW direction. In all maps the H$_2$O emission maximum remains well centered at the position of HH\,211-mm. Higher energy ortho-water transitions such as the $3_{21}-2_{12}$ in the upper left panel of Fig.~\ref{fig:2} display localized patches of emission over a linear path that connects the SE and NW regions while running through the position of the driving source. The same pattern can also be noticed for a series of weaker, tentative H$_2$O detections presented in Fig.~\ref{onlinefig:1} showing also in a couple of cases a locus of relatively bright emission to the NW. 
 
 \begin{figure}[!ht]
\includegraphics[width=8.8cm]{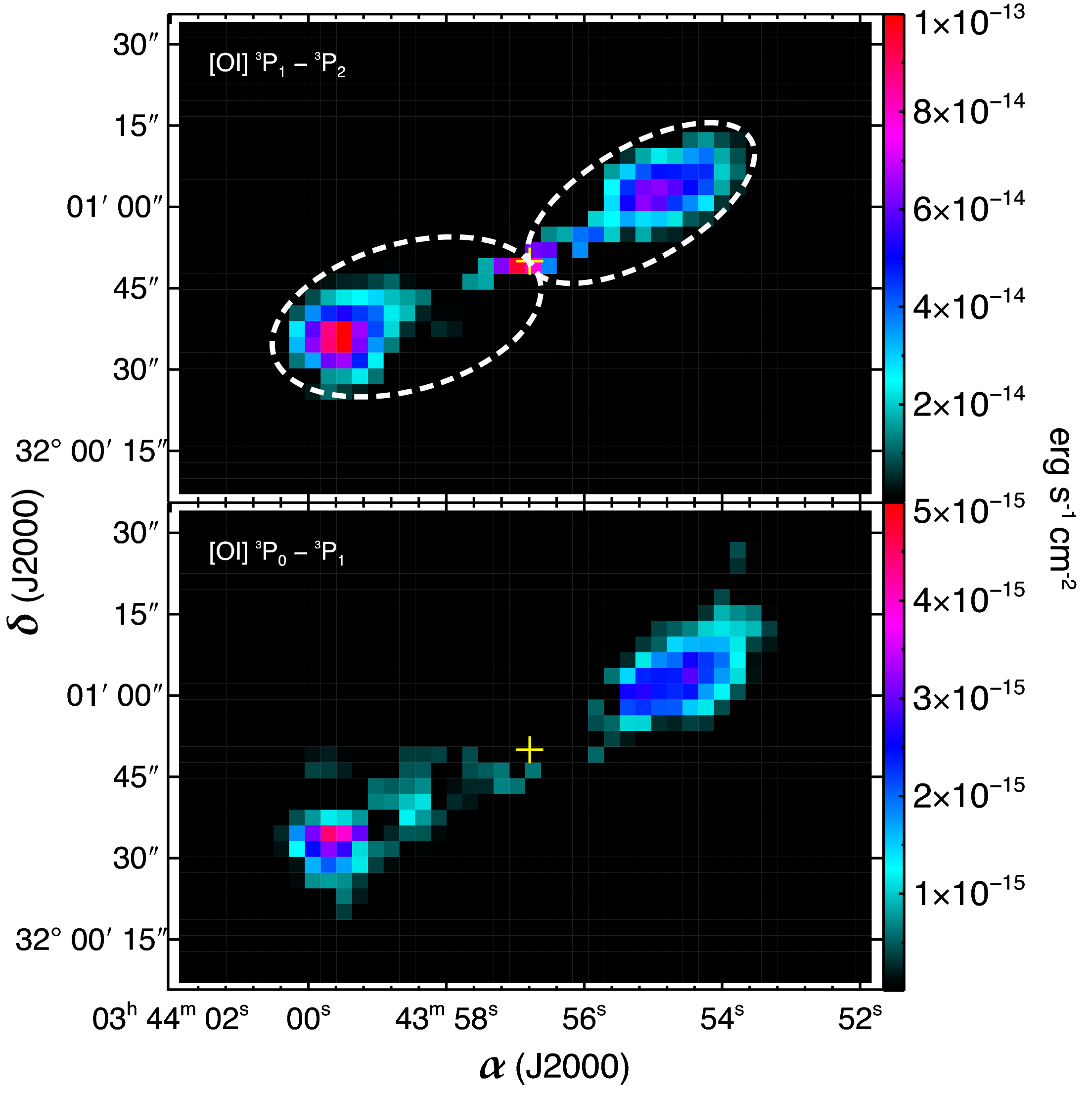}
\caption {Line maps of the $^3P_1$ -- $^3P_2$ and the $^3P_0$ -- $^3P_1$ [\ion{O}{i}] transitions at 63.18 and 145.5~$\mu$m (upper and lower panels, respectively). Notice that color-encoded flux levels are a factor of $\sim$15 higher in the 63.18~$\mu$m map. Dashed ellipses in the upper panel delineate the outflow dimensions considered in the calculation of the [\ion{O}{i}] mass flux (see Sect.~\ref{sec:4.3})} 
\label{fig:4}
\end{figure}
 
In contrast to the emission pattern in the ortho-H$_2$O maps, para-H$_2$O has a very distinct morphology, displaying emission \emph{only} around the position of the protostar. The emission pattern is consistent, being elongated in the brightest para transition and rather rounded in the other two. This apparent difference can be attributed to the line strength and lack of sufficient contrast in the fainter maps.    

The elongated, centrally-peaked emission pattern for both ortho- and para-water forms has only some resemblance to on-source emission of the lowest $J$ CO transitions presented in Fig.~\ref{fig:1}. The major axis of the central emission measures $\sim30\arcsec$, which approximately equals the size of the two-sided SiO jet \citep{Lee:07a}. Based on this similarity some of the water emission may arise from the chain of shocks within the SiO molecular jet, as appears to be the case for the $3_{21}-2_{12}$ transition map. However the peak of the H$_2$O emission is always coincident with the protostar, which indicates that the latter is the main source of excitation for the  H$_2$O lines. In this scenario, the elongated emission pattern to the SE and NW can be attributed to radiation generated very close to the protostar which then escapes through the outflow cavities and excites the material on the cavity walls. Radiative excitation of the envelope material may also explain the detection of para-water lines only in the surroundings of the protostellar source (see Sect.~\ref{sec:4.2}). These findings are also consistent with the 557GHz H$_2$O line-shape observed with Herschel HiFi  \citep{Tafalla:13a}, showing slowly moving material which cannot be associated with the outflows. Based on the morphological similarities between the H$_2$ emission as recorded in the Spitzer/IRAC bands and the $2_{12}-1_{01}$ H$_2$O emission maps, \citet{Tafalla:13a} conclude that the two emissions correlate in both shape and intensity. The higher energy maps presented here demonstrate that the H$_2$O morphology can be significantly different compared to the H$_2$ images. In addition the IRAC images used for comparisons in \citet{Tafalla:13a} display a significant contribution from continuum emission in the region around the protostar. While this emission can mimic  the H$_2$O  line-pattern excited within the envelope, it is not detected in the H$_2$ maps presented in \citetalias{Dionatos:10a}.       
 
The hydroxyl emission shown in Fig.~\ref{fig:3} shows a strong peak at the tip of the SE lobe. This coincides with the spot of the highly excited OH emission detected with Spitzer \citep{Tappe:08a} and therefore the Herschel lines are most likely associated with the lower-energy branch of that emission. Morphologically, the OH peak is very similar to the high-$J$ CO pattern presented in Fig.~\ref{fig:1}. In a few cases OH maps show  peaks also at the on-source and NW positions that resemble more the lower-$J$ CO or the $3_{21}-2_{12}$ H$_2$O emission patterns. Similarly collimated emission is also found in the atomic ([\ion{Si}{ii}], [\ion{Fe}{ii}] and [\ion{S}{i}]) maps presented in \citetalias{Dionatos:10a}. The chain of emission spots along the SE - NW axis, and the single strong emission region at the SE tip, seem to confirm the findings of \citet{Wampfler:13a} that OH is mainly excited in shocks. While the SE peak emission corresponds to the tip of the terminal shock, \citet{Tappe:08a} found that the OH excitation is related to the UV radiation produced at the shock-front. This is also consistent with results of the excitation analysis that follows in Sect.~\ref{sec:4.1}. 
 
 The $^3P_1$ -- $^3P_2$ oxygen line is by more than an order of magnitude the brightest line detected in the surroundings of HH211. Such high levels were also found in previous studies of line emission around embedded protostars \citep[e.g.][]{Green:13a}. The oxygen maps presented in Fig.~\ref{fig:4} show two well delineated, nearly symmetric outflow lobes that appear thinner towards the center of symmetry and come in contact at the position of the protostar. Close to the driving source, [\ion{O}{i}] emission appears, to the available angular resolution, narrow and almost linear resembling the high velocity molecular jets observed in CO and SiO \citep[e.g.][]{Gueth:99a, Hirano:06a, Lee:07a, Lee:11a}  This pattern is still clear, albeit less sharp for the $^3P_0$ -- $^3P_1$ transition (lower panel of Fig.~\ref{fig:4}) that is a factor of $\sim15$ weaker than the $^3P_1$ - $^3P_2$ line. The oxygen emission morphology also appears very similar to that of molecular hydrogen \citepalias{Dionatos:10a}, the only difference being that even mid-IR H$_2$ is not traced very close to the protostar due to extinction from the envelope. A close morphological association between the emission pattern of H$_2$ rotational transitions and the [\ion{O}{i}] lines has also been established in the case of the outflows in NGC\,1333 \citep{Dionatos:17a}. The analysis presented in Sect.~\ref{sec:4.3} provides further evidence of a physical relation between [\ion{O}{i}] and H$_2$.

\section{Analysis and discussion}
 \label{sec:4}

\begin{figure*}[!ht]
\includegraphics[width=17.9cm]{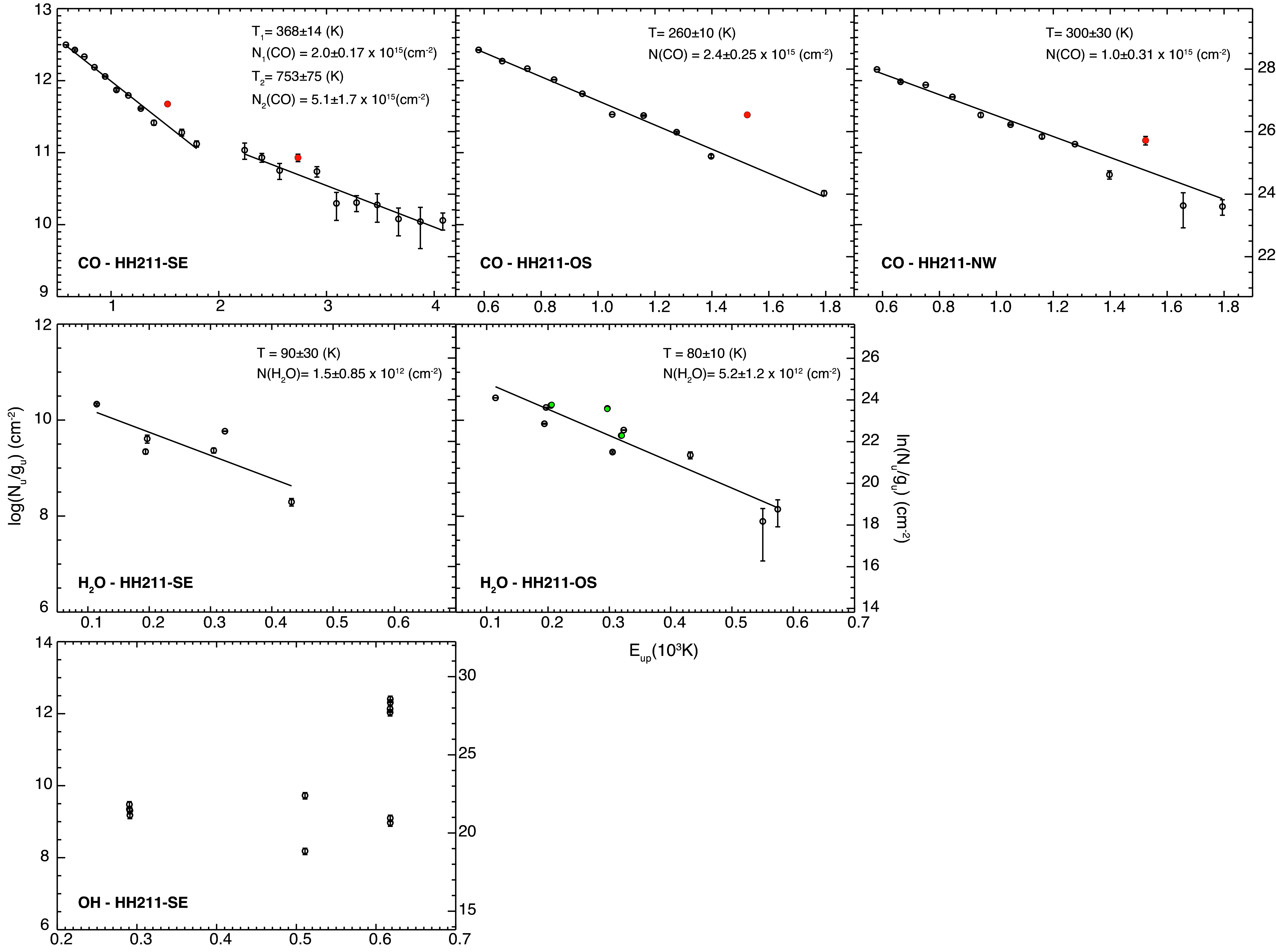}
\caption {Excitation diagrams of CO, H$_2$O and OH (from top to bottom panels) at the southeast (SE), on-source (OS) and northwest positions (NW), respectively (left to right panels). Derived temperatures and column densities are reported in the top-left corner of each panel. Filled (red) circles in the CO panels correspond to blended lines that were not taken into account in the linear fits. Filled (green) circles in the on-source H$_2$O diagram correspond to para-lines.} 
\label{fig:5}
\end{figure*}

\subsection{Excitation} \label{sec:4.1}

To a first approximation we attempt to study the physical conditions of the excited gas by the use of Boltzmann diagrams, in which the populations of an excited level of a molecule are plotted versus the energy of the level involved.
Based on the assumption that the emission is optically thin, the populations of the
upper energy states can be directly derived from line fluxes. 
The slope of the data-point distribution in the diagram is inversely proportional to the excitation
temperature ($T_{\rm ex}$).  The excitation temperature 
of the levels considered is not necessarily the same as the kinetic temperature of the gas, unless the latter is thermalized \citep[e.g.][]{Goldsmith:99a}. The intercept divided by the partition function
provides a measure of the total column density.
Finally, for molecules having ortho and para spin states (e.g. H$_2$ and H$_2$O), the ortho to para ratio can be directly estimated by the relative displacement of the data points corresponding to the two forms \citep[see e.g.][]{Dionatos:13a}.
For the diagrams presented, the upper level energies and
Einstein coefficients were retrieved from the JPL\footnote{http://spec.jpl.nasa.gov/}  and CDMS\footnote{http://www.astro.uni-koeln.de/cdms/}
catalogs \citep[][respectively]{Pickett:98a, Muller:05a}. 
Partition-function tables for a number of distinct temperatures were retrieved from the same databases and the exact values were calculated through interpolation for the corresponding $T_{\rm ex}$. 
In the following we employ excitation diagram diagnostics for the CO, H$_2$O, and OH lines
detected with Herschel/PACS to constrain the physical conditions for each molecule.

The apparent size of the emitting regions in the PACS maps scales from $\sim$\,10$\arcsec$ around 70~$\micron$ to $\sim$\,30$\arcsec$ at 180~$\micron$, reflecting the wavelength dependence of the instrument's angular resolution. Compared to the nominal Herschel resolution at the same wavelengths, the emission is extended and marginally resolved. In order to avoid either flux losses or beam dilution effects that could compromise the excitation analysis, line fluxes were extracted for circular apertures that scale in size linearly with wavelength between 12$\arcsec$ and 33$\arcsec$. Fluxes were extracted for three main regions of emission centered on the source (OS) and the SE and the NW shocks, as shown in the lower-left panel of Fig.~\ref{fig:1}. 

Excitation diagrams of the CO lines for the three positions are presented in the top panels of Fig.~\ref{fig:5}. For the excitation diagram corresponding to the SE region where the highest excitation CO lines are detected, the distribution of data points is best represented by two linear segments with a break around $E_{\rm up}\sim$1800\,K, corresponding to two distinct temperature components. The central and NW positions show no significant excitation of high-$J$ ($J_{\rm up} \geq$ 24) CO lines and can be fit with one line, corresponding to a single temperature. The warm component lies approximately between 300 - 350\,K at the SE and NW positions and a bit lower, at 250K, on source, while the hot component to the SE reaches $\sim$750K.  Estimated column densities for CO range between 10$^{14}$  and 10$^{15}$ cm$^{-2}$ for the warm and hot components which are comparable with previous studies \citep[e.g.][]{Dionatos:13a} and indicate that the main mass carrier of outflows resides the colder component. In a large survey of protostellar sources with Herschel PACS \citep{Karska:18a}, the hot CO component is detected in 24\% of the sources with a median excitation temperature of 720\,K over the whole sample, while the median warm component has a value of $T_{\rm ex}\sim$320\,K. Both these values indicate that the CO temperatures at the shocks are typical for embedded sources, while the on-source position appears somehow cooler. Two temperature components at different positions along the HH211 outflow are also derived from the H$_2$ analysis in \citetalias{Dionatos:10a}. While the warm H$_2$ temperature component traces gas at $T_{\rm ex}\sim$ 300-350\,K like the CO emission, the hot H$_2$ emission peaks at temperatures $\geq$1000\,K indicating higher excitation conditions. 

Excitation diagrams for H$_2$O were created only for the SE and on-source positions where a significant number of water lines were detected. A higher scatter in the distribution of the data points compared to CO is noticeable, which may be attributed to a number of different factors. The water molecule possesses a high critical density which can translate to subthermal excitation, leading to a non-LTE occupation of the different excited levels. At the same time, rotational transitions of water can become optically thick even at rather low densities, which can also increase the apparent scatter in the excitation diagram \citep[see also ][]{vanDishoeck:14a}. Excitation temperature estimates in both the SE and the central positions lie just below 100\,K.  When compared to the median $T_{\rm ex}$ of 130\,K found in \citet{Karska:18a}, the values derived here are somehow lower but still rather representative. 

We plot the excitation diagram for the hydroxyl molecule only for the SE region where enough OH transitions were detected.  Data points are highly scattered and attempts to fit their distribution result in a negative excitation temperature, or at best a very flat distribution that would correspond to excitation temperatures of a few thousand degrees. In the case of OH it becomes obvious that the usability of excitation analysis is limited, as the excitation of the molecule appears to be non-thermal. Spitzer IRS spectra of the SE terminal bowshock, detected a series of highly excited, pure rotational OH lines \citep{Tappe:08a}.  This emission was interpreted as a product of H$_2$O photodissociation that would produce OH molecules in the ground electronic and vibrational states but at high rotational excitation. The rotational levels are then de-excited with a cascade through the OH rotational ladders, and the transitions detected here most likely represent lower-energy steps of the same formation and  de-excitation process.

\subsection{Constraining the CO to H$_2$ ratio in shocks.}

Carbon monoxide is the most abundant molecule after H$_2$ and can be readily excited at very low temperatures which are typical in molecular clouds. Due to these properties and the fact that its lower energy transitions can be easily observed from the ground has rendered CO the most common tracer of protostellar outflows. The determination of mass-related outflow properties relies on the conversion factor between CO and H$_2$, which is commonly assumed to be constant and equal to 10$^{-4}$ \citep{Watson:85a}. The majority of direct determinations of the CO/H$_2$ ratio are based on studies of diffuse clouds and rely on observations of far-ultraviolet absorption lines from the two molecules superimposed on spectra of background stars \citep[e.g.][]{vanDishoeck:87a}. For dense clouds, near-infrared lines in either emission or absorption are employed in order to constrain the abundance ratio \citep[e.g.][]{Lacy:94a}. Rovibrational CO and H$_2$ lines are however excited at temperatures of $\approx$700\,K and $\approx$ 2000\,K, respectively,  and therefore probably trace different volumes of of gas \citep{vanDishoeck:92a}. The method presented below provides a direct determination that has certain advantages compared to previous estimations. 

H$_2$ is the main collisional partner responsible for the excitation of CO in shocks. The excitation analysis confirms that the two molecules are indeed excited in shocks, at essentially the same temperature, and the line maps show that the CO and H$_2$ emissions are also co-spatial. Therefore the rotational CO and H$_2$ lines probe essentially the {\it same} volume of gas. Excitation analysis shows that both molecules are thermalized (i.e. their densities are above the critical density) and therefore their column densities can be well constrained. In addition emission from highly excited CO transitions in shocks is optically thin as they trace high-velocity gas while H$_2$ pure rotational transitions are forbidden and remain optically thin to very high densities. Finally, due to the moderate excitation temperature of  $T_{\rm ex}\sim$ 300-350\,K, no CO dissociation in shocks is expected. 

This line of argument allows us to directly assess the CO abundance by comparing the measured CO and H$_2$ column densities \citep[see also][]{Dionatos:13a}. The median H$_2$ column density for the warm component is 7.65$\times10^{19}$cm$^{-2}$; when compared to the warm CO column densities derived here, we find that $X_{CO}\approx1.3 - 3.0\times10^{-5}$. These values are in excellent agreement with the estimations in \citet{Dionatos:13a} who find a warm-gas $X_{CO}\sim1.8\times10^{-5}$ and provides additional evidence that the galactic CO abundance average of 10$^{-4}$ may not apply globally.

Conditions in shocks are however significantly different when compared to regions of cold, quiescent molecular gas. High temperatures in shocks lead to the sublimation of ices from dust grains which triggers a very rich chemistry where CO and H$_2$ are basic building blocks for a series of more complex molecules (e.g. hydrocarbons). Therefore the CO/H$_2$ ratio measured here provides a more accurate estimation compared to previous direct methods, however because of the particular conditions in shocks, the measured ratio may not directly apply to regions of colder, quiescent gas.

\subsection{Water ortho-to-para ratio}\label{sec:4.2}

Observations of the abundance ratio of nuclear spin isomers of H$_2$O can possibly provide clues about the formation conditions and the thermal processing of molecules. For temperatures above 50\,K the H$_2$O ortho-to-para ratio is $\sim$\,3 while for lower temperatures the ratio tends to zero. Therefore low ortho-to-para (OPR) ratios can be associated with the formation of H$_2$O at very low temperature environments. Processes leading to the desorption of water into the gas phase can alter the low-temperature OPR and modify it toward the thermal equilibrium value of 3. Non-destructive (i.e. maintaining the molecular bonds) photo-desorption of H$_2$O is believed to preserve the original formation value of OPR \citep{vanDishoeck:13a}, however recent laboratory experiments show that photodesorbed water at temperatures as low as 10 K can reproduce a high temperature OPR of 3 which indicates that nuclear spin conversion is active even with low-temperature water-ice \citep{Hama:16a}.

As mentioned in the previous section, both ortho and para forms of water are observed at the position of the protostar.  While there are only three para-H$_2$O lines detected, their distribution in the excitation diagram follows closely the distribution of the ortho-H$_2$O lines (see Fig.~\ref{fig:5}). Attempting a separate excitation analysis on the ortho and para components confirms that the excitation temperature for both forms remains just below 100\,K. The column densities are found to be $\sim$3.9 and 6.1$\times$10$^{12}$ cm$^{-2}$ for the ortho and para forms, respectively. Given the similar excitation conditions, any non-thermal excitation and opacity issues would affect the ortho and para forms in the same way, so that their column density ratio can provide a measure of their abundance ratio. Comparing the measured column densities, we find that an H$_2$O ortho-to-para ratio of $\sim$~0.65. The non-detection of any para-H$_2$O lines in either the SE or NW positions suggests that the ortho-to-para ratio is much higher at the terminal shocks, likely reaching the thermal equilibrium value of three. The same, high temperature equilibrium ratio value at the shock positions has also been measured for the H$_2$ molecule \citepalias{Dionatos:10a}.

The OPR estimated around HH211 is one of the lowest values measured in the surroundings of an embedded protostar, and is within errors comparable to the OPR$\sim$0.77 observed in the outer disk of TW Hya \citep{Hogerheijde:11a}. It is also much lower than expected based on the temperatures of $\sim$90\,K derived from the excitation analysis. This inconsistency between the gas temperature and very low OPR values has been noticed also in the Orion PDR \citep{Choi:14a}. Water around embedded protostars, or even pre-stellar cores, has been proposed to be the product of photodesorption from dust grains in the parental envelope \citep[e.g.][]{Caselli:12a, Mottram:13a}. The action of UV photons in the SE shock that are considered to be responsible for the production of the OH rotational-line cascade (that is partly observed also in the Herschel maps) would be expected to have the same limiting effect on the measured OPR, however this does not appear to be the case. More surprisingly, the same equilibrium OPR is observed on the NW bowshock, where there is no obvious source of UV photons influencing the gas. If the OPR can indeed provide any information about the thermal history of the gas, then the low OPR in the surroundings of the protostar can only be interpreted through processing of primordial material that has never been thermally altered before. At the shock positions in contrast, dust temperature must have been increased multiple times through successive shocks, so that the post-shock formation and subsequent adsorption of water onto/from the dust grains reflects processing at higher temperatures.

\subsection{Mass flux of the [\ion{O}{i}] jet} \label{sec:4.3}

The morphological characteristics of the [\ion{O}{i}] emission presented in Fig.~\ref{fig:4} show a clear jet morphology that impinges the two terminal shocks. To estimate the mass flux of the atomic jet, we use the prescription of \citet{Dionatos:09a}:

\begin{equation}
\dot{M}_{[\ion{O}{i}]} =  \mu m_H \frac{L_{[\ion{O}{i}]}}{h\nu A_if_iX[O]}   \times t_{dyn}^{-1}
\label{eqn:1}
.\end{equation}    

In equation~\ref{eqn:1}, $\mu$ = 1.4 is the mean particle weight per H nucleus, m$_H$ is the mass of the hydrogen atom,  $A_i$ is the Einstein coefficient for spontaneous emission, $f_i$ is the relative occupation of level i, and $X[\rm{O}]$ is the abundance of atomic oxygen. The values for $X[\rm{O}]$ vary from 10$^{-3.52}$ to 10$^{-3.24}$ (Rab et al. 2018, in prep.) and therefore introduce minor uncertainties in the mass calculations.
The mass determination is based on the [\ion{O}{i}] luminosity which is calculated separately for the two outflow lobes summing up the emission from two elliptical regions as shown in Fig.~\ref{fig:4}. The uncertainties related to the [\ion{O}{i}] excitation and the influence of optical depth  and extinction effects are discussed in detail in \citet{Dionatos:17a}. To estimate the mass flux, we defined a dynamical scale from the projected length of the outflows and the tangential velocity of the emitting gas according to the following relation:

\begin{equation}
t_{dyn} = l_t/v_t
\label{eqn:5}
.\end{equation}

Jet-lobe lengths are taken equal to the major axes of the ellipses, measuring 51$\arcsec$ to the SE and 45$\arcsec$ to the NW (see Fig.~\ref{fig:4}). Given the morphological similarities of the SiO jet and the [\ion{O}{i}] close to the protostar, for the tangential velocity we employ v$_t\simeq115$~km\,s$^{-1}$ from the proper motion measurements of the SiO jet \citep{Jhan:16a}. This value does not differ significantly from the assumed jet velocity of 100~km\,s$^{-1}$ employed for the mass flux estimations in \citetalias{Dionatos:10a}, which allows for direct comparisons.
The total mass flux corresponding to the [\ion{O}{i}] emission (i.e. reduced for $X[\rm{O}]$) equals to 8.37$\times 10^{-7}$ and 8.21$\times 10^{-7}$~M$_{\odot}$~yr$^{-1}$ for the SE and the NW lobes, respectively, which sums to 1.65$\times 10^{-6}$~M$_{\odot}$~yr$^{-1}$. 

For an alternative estimation of the [\ion{O}{i}] mass flux, we employ the relation of \citet{Hollenbach:89a} that associates the mass-loss rate and the [\ion{O}{i}] luminosity:

 \begin{equation}
  \frac{\dot{M}_{[\ion{O}{i}]_{shock}}}{ M_{\odot} yr^{-1}}= \frac{10^{-4} \times  L_{[\ion{O}{i}]} }{  L_{\odot}}
\label{eqn:6}
.\end{equation}

which, applied to the total [\ion{O}{i}] luminosity of the lobes yields $3.92\times 10^{-7}$ and $3.57\times 10^{-7}$~M$_{\odot}$ yr$^{-1}$ for the SE and NW lobes, and sums to  $7.5\times 10^{-7}$~M$_{\odot}$ yr$^{-1}$. This value is a factor of two lower compared to the values derived from equation~\ref{eqn:1}. However equation~\ref{eqn:6} is based on the assumption that all the [\ion{O}{i}] emission is generated in a single interface of a dissociative $J-$type shock, which is not consistent with the [\ion{O}{i}] morphology seen in Fig.~\ref{fig:4}. Therefore estimations based on the  \citet{Hollenbach:89a} relation set a lower limit to the total mass-loss rate.
 
The total mass flux based on the emission of  the SiO jet has been estimated to $\dot{M}_{SiO}\sim 0.7 - 2.8 \times 10^{-6}$~M$_{\odot}$~yr$^{-1}$ assuming X[SiO]$\sim$10$^{-6}$ \citep{Lee:07a} or more recently constrained as  $\dot{M}_{SiO}\sim 1.2 \times 10^{-6}$~M$_{\odot}$~yr$^{-1}$ \citep{Jhan:16a}.  Mass flux measurements based on the mid-IR H$_2$ emission maps \citepalias{Dionatos:10a} correspond to $\dot{M}_{H_2}\sim 2 \times 10^{-6}$~M$_{\odot}$~yr$^{-1}$. Therefore mass loss rate estimations for [\ion{O}{i}] are within 20\% in agreement with the mass flux derived from SiO and H$_2$. These results suggest that the atomic jet can dynamically support both the SiO jet and the large-scale H$_2$ emission. We notice however that in contrast to the very similar morphology between H$_2$ and [\ion{O}{i}], the SiO jet is terminated just before the bright terminal bow shocks \citep[e.g.,][]{Hirano:06a} and therefore it possesses a higher momentum density per unit length. The emission of H$_2$ and [\ion{O}{i}] is in contrast dominated by the terminal bow-shocks, which represent the regions where most of the momentum is deposited into the surrounding medium. The mass flux carried by the [\ion{O}{i}] jet is on average a factor of $\sim10$ lower when compared and the mass flux of the CO outflows for a sample of seven protostellar sources in NGC\,1333. The dynamical timescales inferred by equation~\ref{eqn:5} are of the order of just 500 years or less, which is on average a factor of 5 smaller than the average dynamical timescales estimated for the sources in NGC\,1333. Therefore these comparisons suggest that the decline in the [\ion{O}{i}] mass flux reflects an evolutionary trend where mass flux carried out by atomic jets drops as a function of source age, that is similar to the trend observed in the case of molecular outflows \citep{Bontemps:96a}. 

\section{Conclusions} \label{sec:5}

We presented Herschel/PACS spectral mapping observations of the area surrounding the Class 0 source HH\,211-mm.  Molecular lines from CO, H$_2$O and OH and atomic lines from [\ion{O}{i}] were mainly detected in three positions: the two terminal bow-shocks and around the position of the protostellar source. Line maps show some morphological similarities but also pronounced differences. Emission from CO, OH and [\ion{O}{i}] peaks at the tip of the SE bow-shock where Spitzer observations revealed non-thermally excited OH emission. Therefore OH emission at the SE most-likely originates from UV dissociation of H$_2$O at the shock front, which may also be the reason for the weaker water emission in that region. CO excitation analysis at the SE requires two temperature components which are consistent with collisional excitation. 

In striking contrast to all other molecules, water emission peaks at the on-source position where both ortho and para forms are detected. The measured ortho-to-para ratio of just 0.65 indicates formation of water-ice at very low temperatures and a non-destructive photo-desorption process around the protostar.  While part of the H$_2$O emission is likely related to (subthermal) collisional excitation, the centralized morphology around the protostar suggests that radiative excitation is also significant, despite the fact that radiation appears to have a very different impact on the H$_2$O molecules when compared to the SE shock. The very low OPR suggests that H$_2$O around the protostar originates from primordial envelope material that has never been thermally processed before.    

At the NW bowshock emission is on average weaker compared to the SE and on-source positions for all molecules. The non-detection of a hot CO and OH components indicates that the shock conditions in the region are very different compared to the SE shock. Direct measurements of the CO and H$_2$ column densities suggest that the abundance ratio of the two molecules in shocks can be up to an order of magnitude lower than the assumed canonical value of 10$^{-4}$.  To the resolution available, the [\ion{O}{i}] emission appears to trace a well collimated bipolar jet similar to the ones observed with interferometers in CO and SiO in the vicinity of  the protostar, which impacts further out in the surrounding medium in two almost symmetric bowshocks. The [\ion{O}{i}] emission detected with Herschel has close affinity to the H$_2$ line maps observed with Spitzer but is quite different compared to the line emission morphology of CO, H$_2$O and OH. The derived two-sided [\ion{O}{i}] mass flux is 1.65$\times 10^{-6}$~M$_{\odot}$ yr$^{-1}$ that compares well within 20\% to the mas fluxes calculated for the SiO jet and from the H$_2$ emission. This makes HH\,211 the first embedded source where an underlying atomic jet has demonstrably enough power to drive the molecular jet and the larger-scale outflow. 

\begin{acknowledgements}
We would like to thank our referee, who wishes to remain anonymous, for his very constructive comments. This research was supported by the Austrian Research Promotion Agency (FFG)  under the framework of the Austrian Space Applications Program (ASAP) projects JetPro* and PROTEUS (FFG-854025, FFG-866005). T.R. acknowledges support from the ERC Advanced Grant 743029 EASY (Ejection Accretion Structures in YSOs).
\end{acknowledgements}

%
%


\begin{tiny}

\bibliographystyle{aa}
\bibliography{HH211_Herschel}

\end{tiny}


%

\begin{appendix} 
\section{Maps of weaker and tentative H$_2$O line detections.}
\begin{figure}[!ht]
\includegraphics[width=17.9cm]{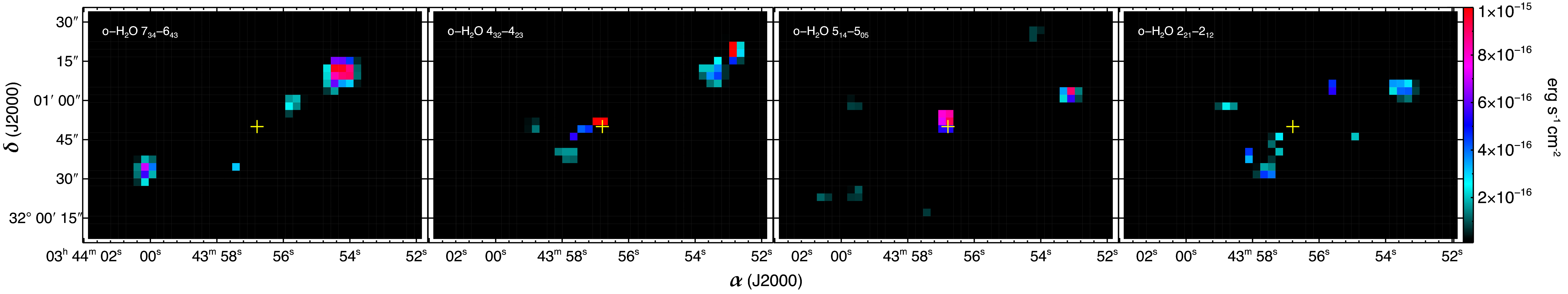}
\caption {Line maps of weaker ortho - H$_2$O lines and marginal detections. } 
\label{onlinefig:1}
\end{figure}

\end{appendix}
\end{document}